\documentclass[11pt]{article}
\usepackage[a4paper,top=80pt,lmargin=80pt,rmargin=80pt,bottom=80pt]{geometry}

\usepackage{graphicx}

\usepackage[]{cite}

\expandafter\let\csname equation*\endcsname\relax
\expandafter\let\csname endequation*\endcsname\relax

\usepackage{amsmath}
\usepackage[]{amssymb}
\usepackage[]{mathrsfs}
\usepackage[]{eucal}
\usepackage[]{tensor}
\usepackage[]{amsthm}
\usepackage[]{bm}

\usepackage{pifont}
\newcommand{\xmark}{\ding{55}}

\newcommand{\pf}[1]{\mathbf{#1}}
\newcommand{\dd}{\partial}
\newcommand{\hdg}{\star} 
\newcommand{\df}{\mathrm{d}}

\newcommand{\nab}[1]{\nabla_{\!#1}}
\newcommand{\qqd}{\ , \quad}
\newcommand{\bc}{\begin{center}}
\newcommand{\ec}{\end{center}}
\newcommand{\be}{\begin{equation}}
\newcommand{\ee}{\end{equation}}

\newcommand{\F}{\pf{F}}

\newcommand{\FF}{\mathcal{F}}
\newcommand{\GG}{\mathcal{G}}

\newcommand{\LL}{\mathscr{L}}

\newcommand{\defeq}{\mathrel{\mathop:}=}

\newcommand{\norm}[1]{\left\lVert #1 \right\rVert}

\usepackage{dsfont}
\newcommand{\nn}{\mathds{N}}

\newcommand{\rr}{\mathds{R}}

\usepackage[unicode]{hyperref}
\usepackage{xcolor}
\definecolor{pastgreen}{HTML}{669900}
\definecolor{pastblue}{HTML}{336699}
\definecolor{pastred}{HTML}{990000}
\definecolor{linkcol}{HTML}{663333}
\hypersetup{colorlinks,linkcolor={pastblue},citecolor={pastblue},urlcolor={pastblue}}

\theoremstyle{plain} \newtheorem{tm}{Theorem}[section]
\theoremstyle{plain} \newtheorem{lm}[tm]{Lemma}
\theoremstyle{plain} \newtheorem{defn}[tm]{Definition}
\newcommand{\btm}{\begin{tm}}
\newcommand{\etm}{\end{tm}}
\newcommand{\blm}{\begin{lm}}
\newcommand{\elm}{\end{lm}}
\newcommand{\bdefn}{\begin{defn}}
\newcommand{\edefn}{\end{defn}}

\newcommand{\arrs}[1]{\renewcommand\arraystretch{#1}}

\begin{document}

\begin{flushright}
\texttt{ZTF-EP-25-04}
\end{flushright}

\vspace{20pt}

\bc
{\huge Weighing the curvature invariants}

\vspace{15pt}

{\Large Jan Draga\v sevi\'c, Ina Moslavac and Ivica Smoli\'c}

\bigskip

e-mails: jan.dragasevic@gmail.com, ina.moslavac@gmail.com, ismolic@phy.hr

\medskip

Department of Physics, Faculty of Science, University of Zagreb, 

Bijeni\v cka cesta 32, 10000 Zagreb, Croatia
\ec

\vspace{15pt}

\begin{abstract}
We prove several inequalities between the curvature invariants, which impose constraints on curvature singularities. Some of the inequalities hold for a family of spacetimes which include static, Friedmann--Lema\^itre--Robertson--Walker, and Bianchi type I metrics, independently of whether they are solutions of some particular field equations. In contrast, others hold for solutions of Einstein's gravitational field equation and a family of energy-momentum tensors (featuring ideal fluids, scalar fields and nonlinear electromagnetic fields), independently of the specific form of the spacetime metric. We illustrate different behaviour of the basic curvature invariants with numerous examples and discuss the consequences and limitations of the proven results.   
\end{abstract}

\vspace{5pt}


\section{Introduction} \label{sec:intro}

Curvature invariants, scalars constructed from the metric $g_{ab}$, Riemann tensor $R_{abcd}$, Levi--Civita tensor $\epsilon_{ab\dots}$ and their covariant derivatives, are ubiquitous in gravi\-tatio\-nal physics \cite{MacCallum15}. They are used as building blocks for Lagrangians, tools for the characterization of spacetimes \cite{CHN09}, regularity conditions in various black hole uniqueness theorems \cite{Heusler} and indicators of singularities \cite{Geroch68,ES77,ES79}.  Nevertheless, half a century after the ``golden age of general relativity'', many questions about the invariants still lurk unanswered.

\smallskip

Construction of scalars from a given set of tensors is a routine combinatorial task, which boils down to permutations of dummy indices. A far more ambitious and difficult challenge is to prove which among the constructed scalars are independent. The fundamental problem of classification of polynomial curvature invariants (without covariant derivative) and a better understanding of their syzygies \cite{Harvey95,Bonanos98,Edgar99} (polynomial relations) has been studied in a series of papers \cite{CMc91,Sneddon96,Sneddon97,Sneddon99,LC04,LC06,LC07}, converging to so-called Zakhary--McIntosh (ZM) set of invariants \cite{ZMc97} (cf.~discussion in \cite{MacCallum15}). Still, even if we have identified algebraically independent invariants, this does not tell us if (and under which conditions) they are ordered by some inequalities.

\smallskip

This question becomes acute in any discussion about the curvature singularities \cite{ES77,ES79}, regions of spacetime in which some of the curvature scalars become ``irregular'' by being either unbounded, wildly oscillating or something else. An immediate dilemma is which scalar(s) exactly to use in such definitions and are these choices somehow related. For example, in Schwarzschild spacetime the Ricci scalar $R$ and the Ricci tensor squared $R_{ab} R^{ab}$ are both (trivially) bounded, whereas the Kretschmann scalar $R_{abcd} R^{abcd}$ is unbounded. Is it possible to have a spacetime in which these scalars behave completely opposite? Of course, emphasis on the Kretschmann scalar stems from its direct relation to tidal forces, but it is easier to calculate the Ricci scalar, so one might ponder if, for example, the unboundedness of the Ricci scalar tells us anything about the other two aforementioned scalars? Our aim is to shed some light on these questions with several inequalities between the invariants and initiate a broader investigation of related properties of spacetimes.

\smallskip

The paper is organized as follows. In section \ref{sec:curv_inv} we introduce the basic curvature invariants and abbreviations, while in section \ref{sec:mean_ineq} we give a brief overview of elementary inequalities which form the backbone of the main results in the paper. Two groups of theorems are split into sections \ref{sec:geom_ineq} and \ref{sec:emt_ineq}: in the former we treat geometric inequalities (valid for a family of spacetimes, independently of whether they are solutions of some particular field equations), while in the latter we treat inequalities for the family of energy-momentum tensors (which hold independently of the specific form of the spacetime metric). In section \ref{sec:examples} we discuss inequalities for various special cases of spacetimes, with examples and counterexamples. Finally, in section \ref{sec:final} we summarize the results with several additional remarks. In Appendix \ref{sec:NLE} we give a brief overview of nonlinear electromagnetic theories, while in Appendix \ref{sec:ZM} we summarize ZM invariants for $(1+3)$-dimensional static, spherically symmetric spacetimes.

\medskip

\emph{Notation}. Throughout the paper, we use the metric signature $(-,+,\dots,+)$ and the natural system of units, in which $G = c = 4\pi\epsilon_0 = 1$. We denote the Minkowski metric with $\eta_{ab}$. For each vector field $v^a$ we use the abbreviation $v^2 \defeq g_{ab} v^a v^b$.

\section{Curvature invariants} \label{sec:curv_inv} 

Unfortunately, there is no universally accepted systematic notation for curvature invariants. Even worse, the same set of symbols in different references are sometimes linear combinations of one another, often without the due caveat (compare e.g.~ZM invariants in the original paper \cite{ZMc97} with those used in \cite{OCWH20,LYGM23}). We shall not attempt to untangle this \emph{confusio linguarum} but instead, keep with the minimal set of auxiliary abbreviations. Three basic scalars in the inventory of curvature invariants are the Ricci scalar $R \defeq g^{ab} R_{ab}$, ``Ricci squared'' $S \defeq R_{ab} R^{ab}$ and the Kretschmann invariant $K \defeq R_{abcd} R^{abcd}$. Furthermore, the Weyl tensor is defined, for $D \ge 3$, as
\be
C_{abcd} \defeq R_{abcd} - \frac{2}{D-2} \, \left ( g_{a[c} R_{d]b} - g_{b[c} R_{d]a} \right ) + \frac{2R}{(D-1)(D-2)} \, g_{a[c} g_{d]b}
\ee
and in $D = 4$ its Hodge dual as
\be
{\hdg C}_{abcd} \defeq \frac{1}{2} \, \epsilon_{abpq} \tensor{C}{^p^q_c_d} \, .
\ee
Traditionally, polynomial curvature invariants are sorted into Weyl invariants (those constructed by contractions of the Weyl tensor and its Hodge dual), Ricci invariants (those constructed by contractions of the Ricci tensor) and mixed invariants (those constructed by a combination of the Ricci tensor, Weyl tensor and its Hodge dual). For example, we have $\mathscr{W}_n = \tensor{C}{^{a_1}^{b_1}_{a_2}_{b_2}} \cdots \tensor{C}{^{a_n}^{b_n}_{a_1}_{b_1}}$ (by construction $\mathscr{W}_1 = 0$) and $\mathscr{R}_n \defeq \tensor{R}{^{a_1}_{a_2}} \tensor{R}{^{a_2}_{a_3}} \cdots \tensor{R}{^{a_n}_{a_1}}$ (e.g.~$\mathscr{R}_1 = R$ and $\mathscr{R}_2 = S$). Note, however, that Zakhary and McIntosh \cite{ZMc97} construct Ricci invariants as contractions $\mathscr{S}_n \defeq \tensor{S}{^{a_1}_{a_2}} \tensor{S}{^{a_2}_{a_3}} \cdots \tensor{S}{^{a_n}_{a_1}}$ of the trace-free Ricci tensor $S_{ab} \defeq R_{ab} - (R/D) g_{ab}$, which are polynomials of the invariants $\mathscr{R}_n$. Furthermore, we have higher derivative invariants, constructed with covariant derivatives, such as $\mathscr{D}_1 = (\nab{a} R)(\nabla^a R)$, $\mathscr{D}_2 = (\nab{a} R_{bc}) (\nabla^a R^{bc})$, $\mathscr{D}_3 = (\nab{a} R_{bcde}) (\nabla^a R^{bcde})$ and so on. Invariant $\mathscr{D}_3$ has caught attention after it was noticed \cite{KLA82} (cf.~discussion in\cite{TK97}) that it vanishes on the horizon of the Schwarzschild black hole, leading to the research of more general horizon-detecting curvature invariants \cite{Gass98,Saa07,CMc18,McMac18}. As far as we are aware, there is no complete, general classification of higher derivative curvature invariants (cf.~\cite{McA86,MacCallum15}) and we shall limit our discussion only to several basic ones.

\section{Mean inequalities} \label{sec:mean_ineq} 

The results studied in this paper rest, among other things, on inequalities between means. A systematic approach starts with the introduction of a generalized mean of positive real numbers $\{x_1,\dots,x_n\}$ for any parameter $p \in \rr^\times$,
\be
M_p(x_1,\dots,x_n) \defeq \left ( \frac{1}{n} \, \sum_{i=1}^n x_i^p \right )^{\frac{1}{p}} .
\ee
In addition, we define $M_0(x_1,\dots,x_n) \defeq (x_1 \cdots x_n)^{1/n}$. For example, $M_{-1}$ is the harmonic mean, $M_0$ is the geometric mean, $M_1$ is the arithmetic mean, $M_2$ is the quadratic mean, etc. Then the mean inequality $M_s(x_1,\dots,x_n) \le M_p(x_1,\dots,x_n)$ holds for any $s < p$ (see e.g. chapter 2 in \cite{HLP}). Since we intend to establish relations involving real numbers which are not necessarily positive, we shall turn to a slight reformulation of the mean inequality, given in the lemma below.

\blm
For any finite nonempty set of real numbers $\{x_1,\dots,x_n\}$ and $p,s\in\nn$, such that $s < p$, the following inequality holds
\be\label{ineq:mean}
\left ( \sum_{i=1}^n |x_i|^s \right )^{\!p} \le n^{p-s} \left ( \sum_{i=1}^n |x_i|^p \right )^{\!s} .
\ee
In particular, if $p$ is an even number, $p = 2m$ for some $m \in \nn$, then the inequality
\be\label{ineq:2m}
\left ( \sum_{i=1}^n x_i^s \right )^{\!2m} \le n^{2m-s} \left ( \sum_{i=1}^n x_i^{2m} \right )^{\!s}
\ee
holds.
\elm

\noindent
\emph{Proof}. A corollary of H\"older's inequality (cf.~\cite{AF}), $\norm{uv}_s \le \norm{u}_p \norm{v}_q$ for $p,q,s \ge 1$ such that $p^{-1} + q^{-1} = s^{-1}$, applied on the set $\{1,\dots,n\}$ with the counting measure, implies
\be
\left ( \sum_{i=1}^n |x_i y_i|^s \right )^{\!\frac{1}{s}} \le \left ( \sum_{i=1}^n |x_i|^p \right )^{\!\frac{1}{p}} \left ( \sum_{i=1}^n |y_i|^q \right )^{\!\frac{1}{q}}
\ee
for any real numbers $x_1$, $y_1$, \dots, $x_n$ and $y_n$. The claim (\ref{ineq:mean}) follows by inserting $y_i = 1$ for all $i$ and raising both sides of the inequality to the power $ps$. Moreover, taking into account that $|\sum_i x_i^s| \le \sum_i |x_i^s| = \sum_i |x_i|^s$ and $|x_i|^{2m} = x_i^{2m}$, the inequality (\ref{ineq:2m}) follows directly from the inequality (\ref{ineq:mean}). \qed

\smallskip

The simplest $m=s=1$ case of (\ref{ineq:2m}) corresponds to the inequality between the arithmetic and the quadratic mean, so-called AM-QM inequality\footnote{We note in passing that the AM-QM inequality also follows from the reorganization of the terms in the elementary inequality $0 \le \sum_{i\ne j} (x_i - x_j)^2$. },
\be
\left ( \sum_{i=1}^n x_i \right )^{\!2} \le n \sum_{i=1}^n x_i^2 \, .
\ee
Utilization of the inequality (\ref{ineq:2m}), and AM-QM inequality in particular, might lead to a suboptimal bound due to several subtle pitfalls \cite{Wilf}. For example, it is certainly true that $(0+x+y)^2 \le 3(0^2 + x^2 + y^2)$ for any $x,y \in \rr$, but it is better to remove irrelevant zeros, if possible, to obtain a stricter bound $(x+y)^2 \le 2(x^2 + y^2)$. Also, in the presence of identical terms, AM-QM inequality implies both $(x+y+y)^2 \leq 3(x^2 + y^2 + y^2)$ and $(x+2y)^2 \leq 2(x^2 + 4 y^2)$, but without further information about the numbers $x$ and $y$, one cannot choose which one is the better bound.

\smallskip

Given a symmetric tensor $X_{ab}$ on a $D$-dimensional pseudo-Riemannian manifold $(M,g_{ab})$, we can construct its trace $X \defeq g^{ab} X_{ab}$ and its trace-free part $X_{ab} - (X/D)\,g_{ab}$. The square of the trace-free part is nonnegative if and only if the inequality $X^2 \le D X^{ab} X_{ab}$ holds. Namely,
\be
\left ( X_{ab} - \frac{X}{D} \, g_{ab} \right ) \left ( X^{ab} - \frac{X}{D} \, g^{ab} \right ) = X^{ab} X_{ab} - \frac{1}{D} \, X^2 .
\ee
This is somewhat reminiscent of an inequality for real $n \times n$ matrices, $(\textrm{Tr}\,A)^2 \le n \textrm{Tr}(A^\mathsf{T}\!A) $, derived with the help of the AM-QM inequality,
\begin{align}
n \textrm{Tr}(A^\mathsf{T}\!A) - (\textrm{Tr}\,A)^2 & = n \sum_{i,j} A_{ij}^2 - \left ( \sum_k A_{kk} \right )^2 \nonumber \\
 & = \left ( n \sum_{i} A_{ii}^2 - \left ( \sum_k A_{kk} \right )^2 \right ) + n\sum_{i \ne j} A_{ij}^2 \ge 0 \, .
\end{align}
Equality above holds if and only if the matrix $A$ is proportional to the unit matrix, $A = \lambda I$, with $\lambda \in \rr$. However, as we are mainly interested in the \emph{Lorentzian} spacetime, such an inequality is nontrivial and demands a careful choice of tensor $X_{ab}$.

\section{Geometric inequalities} \label{sec:geom_ineq} 

We shall first consider relations between the curvature invariants, which are independent of the field equations but hold for specific types of spacetime metrics. In particular, we are focused on three overlapping families of spacetimes, those admitting local coordinates $(t,x^1,\dots,x^{D-1})$ in which the spacetime metric attains one of the forms listed below,
\begin{itemize}
\item[(a)] static,
\be
\df s^2 = -f(x^1,\dots,x^{D-1}) \, \df t^2 + h_{ij}(x^1,\dots,x^{D-1}) \, \df x^i \, \df x^j ,
\ee
with positive $C^2$ function $f$ and $C^2$ Riemannian metric $h_{ij}$;

\smallskip

\item[(b)] generalized FLRW (gFLRW),
\be
\df s^2 = -\df t^2 + a(t)^2 h_{ij}(x^1,\dots,x^{D-1}) \, \df x^i \, \df x^j ,
\ee
with $C^2$ function $a$ and $C^2$ Riemannian metric $h_{ij}$;

\smallskip

\item[(c)] generalized Bianchi type I (gBI),
\be
\df s^2 = -\df t^2 + h_{ij}(t) \, \df x^i \, \df x^j ,
\ee
with $C^2$ Riemannian metric $h_{ij}$.
\end{itemize}

\noindent
We do not assume that metric $h_{ij}$ is either homogeneous or isotropic. Straightforward calculation leads to the following lemma.

\blm\label{lm:RCzero}
For any of the three metrics defined above (static, gFLRW, or gBI) and any indices $i,j,k \in \{1,\dots,D-1\}$ we have $R_{tijk} = 0$ and, consequently, $R_{ti} = 0$ and $C_{tijk} = 0$.
\elm

Now we turn to the first theorem, preceded by a technical lemma about a convenient local coordinate system.

\blm\label{lm:auxcoord}
Let $(M,g_{ab})$ be a $D$-dimensional spacetime, $D \ge 2$, with smooth manifold $M$ and $C^2$ Lorentzian metric $g_{ab}$. Furthermore, suppose that there exists an open subset $U \subseteq M$ with the coordinate chart $(U; t,x^1,\dots,x^{D-1})$, in which the metric has static, gFLRW or gBI form, as described above. Then for each point $p \in U$ there exists a coordinate chart $(V; t',x'^1,\dots,x'^{D-1})$, with $p \in V \subseteq U$, such that (1) the metric attains canonical form at point $p$, $g_{\mu'\nu'}(p) = \eta_{\mu'\nu'}$, and (2) the metric preserves the initial form on the whole set $V$.
\elm

\noindent
\emph{Proof}. Suppose that the coordinates of the point $p$ in the original coordinate chart $(U; t, x^1, \allowbreak \dots, x^{D-1})$ are $(t_p, x_p^1, \dots, x_p^{D-1})$. All three forms of the metric are special cases of the metric of the form
\be
\df s^2 = -f(x^1,\dots,x^{D-1}) \, \df t^2 + h_{ij}(t,x^1,\dots,x^{D-1}) \, \df x^i \, \df x^j . 
\ee
As the metric $g_{ab}$ is, by definition, nondegenerate, we know that $f(x_p^1,\dots,x_p^{D-1}) \ne 0$. Furthermore, let us denote by $\Sigma$ the intersection of the hypersurface $t = t_p$ with the set $U$ and look at the submanifold $(\Sigma,h_{ij})$. We know that there is a neighborhood $W \subseteq \Sigma$ of the point $p \in \Sigma$, and the local coordinate transformation $x'^{i'} = \tensor{A}{^{i'}_j} x^j$ which brings metric components $h_{i'j'}$ into the canonical form at the point $p$. Now, extend this neighborhood to $V = U \cap (\left< t_p - \delta, t_p + \delta \right> \times W)$ and introduce new coordinates via $t' = t \sqrt{|f(p)|}$ and $x'^{i'} = \tensor{A(p)}{^{i'}_j} x^j$. Then
\begin{align*}
g_{t't'} & = \frac{\dd t}{\dd t'} \, \frac{\dd t}{\dd t'} \, g_{tt} + \frac{\dd x^i}{\dd t'} \, \frac{\dd x^j}{\dd t'} \, g_{ij} = -\frac{f(x^1,\dots,x^{D-1})}{|f(x_p^1,\dots,x_p^{D-1})|} , \\
g_{t'i'} & = \frac{\dd t}{\dd t'} \, \frac{\dd t}{\dd x'^{i'}} \, g_{tt} + \frac{\dd x^j}{\dd t'} \, \frac{\dd x^k}{\dd x'^{i'}} \, g_{jk} = 0 , \\
g_{i'j'} & =  \frac{\dd t}{\dd x'^{i'}} \, \frac{\dd t}{\dd x'^{j'}} \, g_{tt} + \frac{\dd x^i}{\dd x'^{i'}} \, \frac{\dd x^j}{\dd x'^{j'}} \, g_{ij} = \tensor{(A(p)^{-1})}{^i_{i'}} \tensor{(A(p)^{-1})}{^j_{j'}} h_{ij} ,
\end{align*}
from where the claim follows. \qed

\medskip

\btm\label{tm:ineqRSK}
Let $(M,g_{ab})$ be a $D$-dimensional spacetime, $D \ge 2$, with smooth manifold $M$ and $C^2$ Lorentzian metric $g_{ab}$. Furthermore, suppose that there exists an open subset $O \subseteq M$ with the coordinate chart $(O; t,x^1,\dots,x^{D-1})$, in which the metric has static, gFLRW, or gBI form, as defined above. Then the following inequalities 
\begin{align}
R^2 & \le DS \label{ineq:RS} \\
2S & \le (D-1)K \label{ineq:SK}
\end{align}
hold at each point of the set $O$.
\etm

\noindent
\emph{Proof}. Let $p \in O$. For all three forms of the metric, we have the auxiliary coordinate chart $(V; t',x'^1,\dots,x'^{D-1})$ from the lemma \ref{lm:auxcoord} at our disposal. As the metric preserves its form on an open set, the vanishing of some component of the Riemann and Ricci tensor in the original coordinate system implies the vanishing of the corresponding component in the auxiliary coordinate system. Thus, lemma \ref{lm:RCzero} implies $R_{t'i'j'k'} = 0$ and $R_{t'i'} = 0$ for all $i',j',k' \in \{1,\dots,D-1\}$. In the rest of the proof, all sums are assumed to go from $1$ to $D-1$. Using the information about the vanishing components at point $p$, we have
\be
R = -R_{t't'} + \sum_{i'} R_{i'i'} \quad \textrm{and} \quad S = (R_{t't'})^2 + \sum_{i',j'} (R_{i'j'})^2 .
\ee
The first inequality (\ref{ineq:RS}) directly follows by the application of the AM-QM inequality. Using the same strategy, we may set upper bounds for different parts of the invariant $S$,
\begin{align*}
 & (R_{t't'})^2 = \left ( \sum_{i'} R_{t'i't'i'} \right )^{\!2} \le (D-1) \sum_{i'} (R_{t'i't'i'})^2 , \\
 & \sum_{i'} (R_{i'i'})^2 = \sum_{i'} \left ( -R_{t'i't'i'} + \sum_{j'\ne i'} R_{j'i'j'i'} \right )^{\!2} \le \\ 
 & \hspace{150pt} \le (D-1) \sum_{i'} \left ( (R_{t'i't'i'})^2 + \sum_{j'\ne i'} (R_{j'i'j'i'})^2 \right ) , \\
 & \sum_{i'} \sum_{j'\ne i'} (R_{i'j'})^2 = \sum_{i'} \sum_{j'\ne i'} \left ( -R_{t'i't'j'} +\!\!\sum_{k'\ne i',j'} \!\!R_{k'i'k'j'} \right )^{\!2} \le \\
 & \hspace{150pt} \le (D-2) \sum_{i'} \sum_{j'\ne i'} \left ( (R_{t'i't'j'})^2 +\!\!\sum_{k'\ne i',j'} \!\!(R_{k'i'k'j'})^2 \right ) ,
\end{align*}
leading to
\begin{align}\label{ineq:2S}
2S \le \sum_{i'} & \left ( 4(D-1) (R_{t'i't'i'})^2 + 2(D-2) \sum_{j'\ne i'} (R_{t'i't'j'})^2 \right ) \, + \nonumber\\
 & + \sum_{i'} \sum_{j'\ne i'} \left ( 2(D-1) (R_{j'i'j'i'})^2 + 2(D-2)\!\!\sum_{k' \ne i',j'} (R_{k'i'k'j'})^2 \right ) .
\end{align}
On the other hand, the Kretschmann invariant $K$ may be decomposed as
\begin{align}
K = 4\sum_{i'} & \left ( (R_{t'i't'i'})^2 + \sum_{j' \ne i'} (R_{t'i't'j'})^2 \right ) + \nonumber\\
 & + \sum_{i'} \sum_{j'\ne i'} \left ( 2(R_{i'j'i'j'})^2 + 4\!\sum_{k'\ne i',j'} \!(R_{i'k'j'k'})^2 \right ) + \sum_{\dots} (R_{i'j'k'\ell'})^2 ,
\end{align}
where the last term denotes the sum over all pairwise distinct indices $(i',j',k',\ell')$. If we multiply the Kretchmann invariant by $(D-1)$, and compare it with terms on the right-hand side of the inequality (\ref{ineq:2S}), we see that the inequality (\ref{ineq:SK}) is indeed satisfied. \qed

\medskip

Given that spacetime metric satisfies conditions of the theorem \ref{tm:ineqRSK}, inequalities (\ref{ineq:RS}) and (\ref{ineq:SK}) imply that $S$ and $K$ are nonnegative on the set $O$ and we have two immediate consequences,

\begin{itemize}
\item[(1)] ``$R \to S \to K$ rule'': if the Ricci scalar $R$ is unbounded along some path through the set $O$, so are $S$ and $K$, and if $S$ is unbounded along some path through the set $O$, so is $K$ (in the latter case we cannot \emph{a priori} say more about $R$);

\item[(2)] ``$K \to S \to R$ rule'': if the Kretschmann scalar $K$ goes to zero along some path through the set $O$, so do $S$ and $R$, and if $S$ goes to zero along some path through the set $O$, so does $R$ (in the latter case we cannot \emph{a priori} say more about $K$).
\end{itemize}

\noindent
As stated, theorem \ref{tm:ineqRSK} holds in the interior of static black holes, as long as $f > 0$ (an example is the region bounded by the inner horizon of the subextremal Reissner--Nordstr\"om black hole). On the other hand, if $f < 0$, metric $h_{ij}$ becomes Lorentzian and invariants $S$ and $K$ in general lose manifestly nonnegative form. Nevertheless, inequalities (\ref{ineq:RS}) and (\ref{ineq:SK}) still hold for some special subcases of static spacetimes, such as spherically symmetric static ones, independently of the sign of the function $f$, as will be shown in the subsection \ref{sec:static}.

\smallskip

It is instructive to look more closely into the low-dimensional cases. In the $(1+2)$-dimensional case (cf.~equation (6.7.6) in \cite{Weinberg} and equation (1.2) in \cite{GD}) the Riemann tensor is completely defined by the Ricci tensor,
\be
R_{abcd} = 2\left ( g_{a[c} R_{d]b} - g_{b[c} R_{d]a} \right ) - R \, g_{a[c} g_{d]b} \, ,
\ee
so that
\be
4S = K + R^2 .
\ee
Note that (a) inequality $R^2 \le 3S$ implies $S \le K$ and, vice versa, (b) inequality $S \le K$ implies $R^2 \le 3S$. Hence, in a $(1+2)$-dimensional case one inequality implies the other via identity $4S = K + R^2$. In the $(1+1)$-dimensional case (cf.~equation (6.7.4) in \cite{Weinberg}) the Riemann tensor is completely defined by the Ricci scalar,
\be
R_{abcd} = R \, g_{a[c} g_{d]b} \, ,
\ee
so that
\be
R^2 = 2S = K \, .
\ee
Hence, in the $(1+1)$-dimensional case both inequalities are reduced to equalities.

\smallskip

Let us turn to other curvature invariants for $D \ge 3$. We shall denote the simplest nontrivial ``Weyl squared'' with $W \defeq \mathscr{W}_2 = C_{abcd} C^{abcd}$. We know that in general $W$ is not positive definite \cite{Schmidt03}. However, for any metric covered by the theorem \ref{tm:ineqRSK}, using the same decomposition in the coordinate chart $(O; t,x^1,\dots,x^{D-1})$ as for the Kretschmann invariant, it follows that $W \ge 0$. An identity connecting invariants $R$, $S$, $K$ and $W$,
\be\label{eq:RSKW}
K + \frac{2R^2}{(D-1)(D-2)} = W + \frac{4S}{D-2} \, ,
\ee
can be obtained by straightforward, but tedious calculation. It is convenient to rewrite this identity in the form
\be\label{eq:RSKWb}
\big( (D-1)K - 2S \big) - \frac{2}{D-2} \, \left ( DS-R^2 \right ) = (D-1)W
\ee
in order to reveal the relation between $W$ and differences $(D-1)K - 2S$ and $DS-R^2$, controlled by inequalities (\ref{ineq:RS}) and (\ref{ineq:SK}). There are two exceptional cases which should be stressed: for $D=3$, as well as $D=4$ Petrov type O metrics, the Weyl tensor vanishes identically, $C_{abcd} = 0$, implying $W = 0$ and $(D-2)\left ( (D-1) K - 2S \right ) = 2(DS - R^2)$.

\smallskip

Many of the Weyl and mixed invariants in $(1+3)$-dimensional spacetimes constructed with the Hodge dual ${\hdg C}_{abcd}$ are identically zero for any metric covered by the theorem \ref{tm:ineqRSK}. Take, for example, $2\tensor{C}{^{\mu_1}^{\nu_1}_{\mu_2}_{\nu_2}} \tensor{{\hdg C}}{^{\mu_2}^{\nu_2}_{\mu_1}_{\nu_1}} = \epsilon_{\alpha\beta\gamma\delta} \tensor{C}{^\alpha^\beta_\sigma_\tau} C^{\gamma\delta\sigma\tau}$ in the coordinate chart $(O; t,x^1,\dots,x^{D-1})$. For any nontrivial term at least one of the Levi--Civita indices has to be $t$; let us, without loss of generality, assume that $\alpha = t$. However, as $C_{tijk} = 0$, we know that one of the indices $\sigma$ or $\tau$ has to be equal to $t$, and the same reasoning applies to indices $\gamma$ and $\delta$ leading to the conclusion that $\tensor{C}{^{a_1}^{b_1}_{a_2}_{b_2}} \tensor{{\hdg C}}{^{a_2}^{b_2}_{a_1}_{b_1}} = 0$. The analogous argument may be applied to the ZM invariants denoted by $I_4$, $I_{10}$, $I_{12}$ and $I_{17}$ in \cite{OCWH20} (in fact, all of them are proportional to the spin parameter $a$ in the case of Kerr--Newman spacetime, which reduces to the static Reissner--Nordstr\"om spacetime in the $a \to 0$ limit). This reasoning is reminiscent of the one used for the proof of the vanishing of the family of gravitational Chern--Simons terms for metrics of the specific block-diagonal form \cite{BCDPPS11b,BCDPPS13}. Further study of the relations between the rest of the polynomial curvature invariants is left for future work.

\smallskip

Inequalities for the higher derivative invariants rely on more delicate assumptions. More concretely, as shown in the following theorem and discussed in section \ref{sec:examples}, inequalities between $\mathscr{D}_1$, $\mathscr{D}_2$, and $\mathscr{D}_3$ hold for a metric with static form, given that the vector field $\dd_t$ is timelike (this assumption cannot be relaxed even in the spehrically symmetric case).

\btm\label{tm:ineqD}
Let $(M,g_{ab})$ be a spacetime with a metric which on a coordinate chart $(O; t,x^1,\dots,x^D)$ has the static form with $f > 0$ on $O$. Then
\begin{align}
\mathscr{D}_1 & \le D \mathscr{D}_2 \label{D1D2} \\
2\mathscr{D}_2 & \le (D-1)\mathscr{D}_3 \label{D2D3}
\end{align}
\etm

\noindent
\emph{Proof}. Using decompositions
\begin{align*}
\mathscr{D}_1 & = \sum_{i'} \left ( \nab{i'} R \right )^2 = \sum_{i'} \left ( {-\nab{i'}} R_{t't'} + \sum_{j'} \nab{i'} R_{j'j'} \right )^{\!2} \, , \\
\mathscr{D}_2 & = \sum_{i'} \left ( 2(\nab{t'} R_{t'i'})^2 + (\nab{i'} R_{t't'})^2 \right ) + \sum_{i',j',k'} \left ( \nab{i'} R_{j'k'} \right )^2 \, ,
\end{align*}
AM-QM inequality implies
\be
\mathscr{D}_1 \le D \sum_{i'} \left ( \left ( \nab{i'} R_{t't'} \right )^2 + \sum_{j'} \left ( \nab{i'} R_{j'j'} \right )^2 \right ) \le D \mathscr{D}_2 \, .
\ee
Furthermore, using decompositions
\begin{align*}
\mathscr{D}_2 & = \sum_{i'} \left ( 2 \left ( \sum_{j'\ne i'} \nab{t'} R_{t'j'i'j'} \right )^{\!2} + \left ( \sum_{j'} \nab{i'} R_{t'j't'j'} \right )^{\!2} \right ) \, + \\
 & \hspace{100pt} + \sum_{i',j',k'} \left (- \nab{i'} R_{t'j't'k'} + \sum_{\ell'\ne j',k'} \nab{i'} R_{j'\ell'k'\ell'} \right )^{\!2} \, , \\
\mathscr{D}_3 & = 4 \sum_{i',j',k'} \left ( \left ( \nab{t'} R_{t'i'j'k'} \right )^2 + \left ( \nab{i'} R_{t'j't'k'} \right )^2 \right ) + \sum_{\dots} \left ( \nab{i'} R_{j'k'\ell'm'} \right )^{\!2} \, ,
\end{align*}
AM-QM inequality implies (note that for $j' = k'$ the $\ell'$ sum has $D-2$ terms, whereas for $j' \ne k'$ it has $D-3$ terms)
\begin{align}
2\mathscr{D}_2 & \le \sum_{i'} \left ( 4(D-2) \sum_{j'\ne i'}  \left ( \nab{t'} R_{t'j'i'j'} \right )^2 + 2(D-1) \sum_{j'} \left ( \nab{i'} R_{t'j't'j'} \right )^2 \right ) \, + \nonumber \\
 & \hspace{100pt} + \sum_{i',j',k'} 2(D-1) \left ( \left ( \nab{i'} R_{t'j't'k'} \right )^2 + \sum_{\ell'\ne j',k'} \left ( \nab{i'} R_{j'\ell'k'\ell'} \right )^2 \right ) \nonumber \\
 & \le (D-1) \mathscr{D}_3 \, ,
\end{align}
which proves the claim. \qed

\section{Energy-momentum tensor inequalities} \label{sec:emt_ineq} 

Now we turn to solutions of Einstein's gravitational field equation, 
\be
R_{ab} - \frac{1}{2} \, R g_{ab} = 8\pi T_{ab} \, ,
\ee
without \emph{a priori} assumptions about the form of the metric $g_{ab}$. Here we have a direct relation between the energy-momentum tensor $T_{ab}$ and its trace $T \defeq g^{ab} T_{ab}$ on one side, and Ricci tensor $R_{ab}$ and Ricci scalar $R$ on the other,
\be
R_{ab} = 8\pi \left( T_{ab} - \frac{T}{D-2} \, g_{ab} \right) , \quad R = -\frac{16\pi}{D-2} \, T .
\ee
Hence, one can directly translate relations between scalars constructed from the energy-momentum tensor into relations between Ricci invariants. For example,
\be\label{eq:SRTTT}
DS - R^2 = (8\pi)^2 (D T^{ab} T_{ab} - T^2) .
\ee
We first bring into focus a family of energy-momentum tensors described in the theorem below.

\btm\label{tm:Tineq}
Let $(M,g_{ab})$ be $D$-dimensional Lorentzian manifold, on which we have defined
\begin{itemize}
\item[(a)] $1 \le N \le D$ vector fields $\{ v^a_{(1)}, \dots, v^a_{(N)} \}$,

\item[(b)] $1 \le N \le D$ real functions $\{f_1, \dots, f_N \}$, $f_i : M \to \rr$, such that all $f_i \ge 0$ if $N \ge 2$, and

\item[(c)] real function $F : M \to \rr$.
\end{itemize}

\noindent
Then the tensor
\be 
T_{ab} = \sum_{i=1}^N f_i v_a^{(i)} v_b^{(i)} + F g_{ab}
\ee 
is symmetric and the square of its trace-free part is nonnegative, 
\be\label{ineq:T}
T^2 \le D T^{ab} T_{ab} \, .
\ee
\etm

\noindent
\emph{Proof}. By straightforward calculation, we have
\begin{align*}
D T^{ab} T_{ab} - T^2 & = D \sum_i f_i^2 (v_{(i)}^2)^2 + D \sum_i \sum_{j\ne i} f_i f_j (v_{(i)}^a v^{(j)}_a)^2 \, + \\
 & \hspace{30pt} + 2DF \sum_i f_i v_{(i)}^2 + (DF)^2 - \left ( \sum_i f_i v_{(i)}^2 + DF \right )^2 \\
 & = D \sum_i f_i^2 (v_{(i)}^2)^2 - \left ( \sum_i f_i v_{(i)}^2 \right )^2 + D \sum_i \sum_{j\ne i} f_i f_j (v_{(i)}^a v^{(j)}_a)^2 \\
 & \ge N \sum_i f_i^2 (v_{(i)}^2)^2 - \left ( \sum_i f_i v_{(i)}^2 \right )^2 + D \sum_i \sum_{j\ne i} f_i f_j (v_{(i)}^a v^{(j)}_a)^2 ,
\end{align*}
where all sums go from $1$ to $N$. Application of the AM-QM inequality implies that the difference between the first two terms is nonnegative, while assumptions about functions $f_i$ guarantee that the third term is nonnegative. \qed

\medskip

The physically most interesting application of the result above is to the energy-momentum tensor $T_{ab}$ on $(1+3)$-dimensional spacetime. Energy-momentum tensors which belong to the family covered by the theorem \ref{tm:Tineq} include those for
\begin{itemize}
\item[(1)] ideal fluid,
\be
T_{ab} = (\rho + p) u_a u_b + p g_{ab}
\ee
with density $\rho$, pressure $p$ and fluid element 4-velocity $u^a$ (formally, the cosmological constant term is an ideal fluid with $\rho = -p = \Lambda/(8\pi)$), as well as anisotropic fluid with the energy-momentum tensor of the form
\be
T_{ab} = (\rho + p_\perp) u_a u_b + p_\perp g_{ab} - (p_\perp - p_\parallel) w_a w_b \, ,
\ee
with unit spacelike vector field $w^a$ orthogonal to $u^a$ and pressure components, $p_\perp$ and $p_\parallel$, perpendicular and parallel to $w^a$;

\smallskip

\item[(2)] real scalar field,
\be
T_{ab} = \nab{a}\phi \nab{b}\phi + \left ( X - \mathscr{U}(\phi) \right ) g_{ab}
\ee
with the abbreviation $X \defeq -(\nab{a}\phi \nabla^a\phi)/2$ and potential $\mathscr{U}(\phi)$, as well as ``k-essence'' generalization
\be
T_{ab} = p_{,X} \nab{a}\phi \nab{b}\phi + p g_{ab} \, ,
\ee
with $p = p(\phi,X)$;

\smallskip

\item[(3)] complex scalar field,
\be
T_{ab} = \nab{(a}\phi^* \nab{b)}\phi - \frac{1}{2} \, \left ( \nab{c}\phi^* \nabla^c\phi + \mathscr{U}(\phi^*\phi) \right ) g_{ab} \, ,
\ee
with the remark that real-imaginary decomposition of the field $\phi = \alpha + i\beta$ implies $\nab{(a}\phi^* \nab{b)}\phi = \nab{a}\alpha \nab{b}\alpha + \nab{a}\beta \nab{b}\beta$.
\end{itemize}

\noindent
It is important to stress that for a single-component energy-momentum tensor the inequality (\ref{ineq:T}) is independent of the energy conditions: e.g.~it holds for ideal fluid with any density $\rho$ and pressure $p$, whereas energy conditions set some constraints (cf.~chapter 9 in \cite{Wald}) on $\rho$ and $p$. On the other hand, if we have a combination of the energy-momentum tensors above, nonnegativity conditions $f_i \ge 0$ from the theorem \ref{tm:Tineq} impose additional constraints. In the case when the matter consists of several components of the ideal fluid, each with density $\rho_i$ and pressure $p_i$, we need to assume that $\rho_i + p_i \ge 0$ for all $i$ (this condition holds if, for example, each component of the fluid satisfies the weak energy condition \cite{Wald}) to assure the inequality (\ref{ineq:T}).

\smallskip

Let us look at the higher order contractions of the energy-momentum tensor, $\mathscr{T}_p \defeq \tensor{T}{^{a_1}_{a_2}} \tensor{T}{^{a_2}_{a_3}} \cdots \tensor{T}{^{a_p}_{a_1}}$. We shall, for simplicity, take a single-component energy-momentum tensor of the form $T_{ab} = f v_a v_b + F g_{ab}$, for which it can be shown that $\mathscr{T}_p = (f v^2 + F)^p + (D-1) F^p$. We can insert $x_1 = fv^2 + F$ and $x_2 = \cdots = x_D = F$ in (\ref{ineq:2m}) with $n=D$, in order to get a family of inequalities
\be
(\mathscr{T}_s)^{2m} \le D^{2m-s} (\mathscr{T}_{2m})^s
\ee
for all integers $s$ and $m$, such that $1 \le s < 2m$. For example, $s=1$ gives us a generalization of the inequality (\ref{ineq:T}),
\be
T^{2m} \le D^{2m-1} \mathscr{T}_{2m} \, .
\ee
Another way to reformulate the result is to introduce a trace-free part of the energy-momentum tensor
\be
\widehat{T}_{ab} \defeq T_{ab} - \frac{T}{D} \, g_{ab}
\ee
and the corresponding contractions $\widehat{\mathscr{T}}_p \defeq \tensor{\widehat{T}}{^{a_1}_{a_2}} \tensor{\widehat{T}}{^{a_2}_{a_3}} \cdots \tensor{\widehat{T}}{^{a_p}_{a_1}}$. For the particular case of $T_{ab} = f v_a v_b + F g_{ab}$ we have $\widehat{T}_{ab} = f v_a v_b - (f v^2/D) g_{ab}$, essentially just a replacement of $F$ with $-f v^2/D$, implying immediately
\be\label{ineq:hatT}
\left ( \widehat{\mathscr{T}}_s \right )^{2m} \le D^{2m-s} \left ( \widehat{\mathscr{T}}_{2m} \right )^s .
\ee

Suppose now that the spacetime metric $g_{ab}$ is a solution of Einstein's gravitational field equation $G_{ab} = 8\pi T_{ab}$. Then the inequality $T^2 \le D T_{ab} T^{ab}$, via relation (\ref{eq:SRTTT}), implies inequality $R^2 \le DS$. As for the higher-order Ricci invariants, the most straightforward relation stems from the proportionality of trace-free parts of the energy-momentum and Ricci tensor,
\be
S_{ab} = R_{ab} - \frac{R}{D} \, g_{ab} = 8\pi \left( T_{ab} - \frac{T}{D} \, g_{ab} \right) = 8\pi \widehat{T}_{ab}
\ee
leading to $\mathscr{S}_n = (8\pi)^n \widehat{\mathscr{T}}_n$ and all the associated inequalities. The gravitational field equation does not offer us a direct insight into the complete Riemann tensor, so in general we cannot establish a relation between the contractions $\mathscr{T}_n$ and e.g.~Kretschmann invariant. However, if we assume that the metric $g_{ab}$ is a solution of the Einstein's field equation, with the energy-momentum tensor belonging to family covered by the theorem \ref{tm:Tineq}, \emph{and} that metric belongs to one of the families of spacetimes covered by the theorem \ref{tm:ineqRSK}, then for $D \ge 3$ we have an inequality
\be
\frac{(8\pi)^2}{D} \, T^2 \le (8\pi)^2 \, T_{ab} T^{ab} = S + \frac{D-4}{4}\,R^2 \le \frac{(D-1)(D-2)^2}{8} \, K.
\ee
For example, $D=4$ ideal fluid under given assumptions satisfies $32\pi^2 (3p-\rho)^2 \le 3K$.

\smallskip

Finally, we can extend the results above to the energy-momentum tensor of the nonlinear electromagnetic (NLE) field (see Appendix A). We shall, for simplicity, focus on $(1+3)$-dimensional spacetimes, with the NLE Lagrangian density $\LL(\FF,\GG)$, where $\FF \defeq F_{ab} F^{ab}$ and $\GG \defeq F_{ab} {\hdg F}^{ab}$ are electromagnetic invariants.

\btm\label{tm:TNLE}
At each point of the $(1+3)$-dimensional spacetime, contractions of the trace-free part of the NLE energy-momentum tensor satisfy
\be
\widehat{\mathscr{T}}_{2n+1} = 0 \qqd \widehat{\mathscr{T}}_{2n} = 4 \, \frac{(\LL_\FF)^{2n}}{(4\pi)^{2n}} \, (\FF^2 + \GG^2)^n
\ee
for all $n \in \nn_0$.
\etm

\noindent
\emph{Proof}. Higher order contractions of the Maxwell's electromagnetic energy-momentum tensor $T^{(\textrm{Max})}_{ab}$ have a particularly simple form \cite{BSJ22b},
\be
\mathscr{T}^{(\textrm{Max})}_{2n+1} = 0 \qqd \mathscr{T}^{(\textrm{Max})}_{2n} = \frac{4}{(16\pi)^{2n}} \, (\FF^2 + \GG^2)^n .
\ee
Hence, the claim in the theorem directly follows if these expressions are used in contractions of the trace-free part of the NLE energy-momentum tensor (\ref{eq:trfreeNLE}). \qed

\medskip

An immediate consequence is that for NLE fields all these contractions are nonnegative, $0 \le \widehat{\mathscr{T}}_{2n}$, and interrelated via $4^{2n} (\widehat{\mathscr{T}}_{2n})^{2m} = 4^{2m} (\widehat{\mathscr{T}}_{2m})^{2n}$. In particular, we have
\be
0 \le \widehat{\mathscr{T}}_2 = T^{ab} T_{ab} - \frac{1}{4} \, T^2 ,
\ee
where the strict inequality $T^2 < 4 T^{ab} T_{ab}$ holds if $\LL_\FF \ne 0$ and the electromagnetic field is not null.

\section{Examples and counterexamples} \label{sec:examples} 

The most important stationary, axially symmetric electrovacuum solution of the Einstein--Maxwell field equations is the Kerr--Newman spacetime, representing a charged rotating black hole with the mass $M$, spin parameter $a$, and electric charge $Q$. As a direct consequence of the tracelessness of Maxwell's electromagnetic energy-momentum tensor, $T = 0$, we know that $R = 0$ for this solution. Moreover, we have\footnote{All ZM invariants for the Kerr--Newmann spacetime were explicitly calculated in \cite{OCWH20}.},
\begin{align}
S & = \frac{4Q^4}{(r^2 + a^2\cos^2\theta)^4} \, , \\
K & = \frac{8}{(r^2 + a^2\cos^2\theta)^6} \, \Big ( (7Q^4 - 12MQ^2 r + 6M^2 r^2) r^4 \, - \nonumber\\
 & \hspace{30pt} - 2a^2 (17 Q^4 - 60M Q^2 r + 45M^2 r^2) r^2 \cos^2\theta \, + \nonumber\\
 & \hspace{30pt} + a^4 (7Q^4 - 60M Q^2 r + 90M^2 r^2) \cos^4\theta - 6 a^6 M^2 \cos^6\theta \Big ) .
\end{align}
Invariant $S$ is nonnegative, in accordance with the theorem \ref{tm:TNLE} and the fact that $S = (8\pi)^2 \mathscr{T}_2^{\mathrm{(Max)}}$, while none of the results above put any constraint on the difference $3K - 2S$. For example, for the Kerr black hole ($Q = 0$) on the axis of symmetry ($\theta = 0,\pi$) we have
\be
3K - 2S = \frac{144 M^2 P(r)}{(r^2 + a^2)^6} \, ,
\ee
with the polynomial $P(r) = (r^2 - a^2)(r^4 - 14a^2 r^2 + a^4)$, whose positive zeros are $r_1 = (2-\sqrt{3})a$, $r_2 = a$ and $r_3 = (2+\sqrt{3})a$. Polynomial $P$, and thus the difference $3K - 2S$ along the axis, is negative on $\left< 0,r_1 \right> \cup \left< r_2,r_3 \right>$ and positive on $\left< r_1,r_2 \right> \cup \left< r_3,\infty \right>$. Let us now look more closely into representative examples of spacetimes covered by the theorem \ref{tm:ineqRSK}.

\subsection{Static, spherically symmetric spacetimes} \label{sec:static} 

The most researched static spacetimes are spherically symmetric ones. More concretely, we shall assume that $D \ge 3$ and that the metric satisfies a special condition $g_{tt} g_{rr} = -1$ (cf.~\cite{Jacobson07}),
\be
\df s^2 = -f(r) \, \df t^2 + \frac{\df r^2}{f(r)} + r^2 \df\Omega_{D-2}^2 \, ,
\ee
where $\df\Omega_{D-2}^2$ is the round metric on unit $(D-2)$-sphere. The auxiliary coordinate system from lemma \ref{lm:auxcoord}, on a connected neighborhood of the point $(t_0,r_0,\dots)$ on which $f \ne 0$, may be constructed via
\be
t' = t\,\sqrt{|f(r_0)|} \qqd r' = r_0 + \int_{r_0}^r \frac{\df s}{\sqrt{|f(s)|}} \, ,
\ee
and $\theta'_i = \theta_i$ for all the angular coordinates.

\smallskip

Corresponding basic curvature invariants have a relatively simple form,
\begin{align}
- R & = f'' + 2 \, \frac{D-2}{r} \, f' + \frac{(D-2)(D-3)}{r^2} \, (f-1) , \\
S & = \frac{1}{2} \left( f'' + \frac{D-2}{r} \, f' \right)^{\!2} + (D-2) \left( \frac{f'}{r} + \frac{D-3}{r^2} \, (f-1) \right)^{\!2} , \\
K & = (f'')^2 + 2 \, \frac{D-2}{r^2} \, (f')^2 + 2 \, \frac{(D-2)(D-3)}{r^4} \, (f-1)^2 , \\
W & = \frac{D-3}{D-1} \, \frac{\left ( r^2 f'' - 2r f' + 2(f - 1) \right )^2}{r^4} \, .
\end{align}
We can gain insight into the inequalities (\ref{ineq:RS}) and (\ref{ineq:SK}) from the theorem \ref{tm:ineqRSK}, by writing differences between the left and right-hand side as manifestly nonnegative functions, 
\begin{align}
DS - R^2 & = \frac{D-2}{2} \left( f'' + \frac{D-4}{r} f' - \frac{2(D-3)}{r^2} \, (f-1) \right)^{\!2}  , \\
(D-1) K - 2S & = (D-2) \left ( \left ( f'' - \frac{f'}{r} \right )^{\!2} + (D-3) \left ( \frac{2}{r^2} \, (f-1) - \frac{f'}{r} \right )^{\!2} \right ) .
\end{align}
Note that $DS - R^2$ and $(D-1) K - 2S$ are nonnegative for any sign of the function $f$. In the $D = 3$ case, the equalities $3S = R^2$ and $K = S$ hold on an nonempty open set iff $f(r) = c_1 + c_2 r^2$ with constants $c_1$ and $c_2$. On the other hand, in the $D \ge 4$ cases equality $DS = R^2$ holds on an nonempty open set iff $f(r) = 1 + \alpha r^2 + \beta r^{3-D}$ with constants $\alpha$ and $\beta$, which is isometric to a subset of $D$-dimensional Schwarzschild-(anti)-de Sitter spacetime, whereas equality $(D-1)K = 2S$ holds on an nonempty open set iff $f(r) = 1 + \gamma r^2$ with constant $\gamma$, which is isometric to a subset of $D$-dimensional (anti)-de Sitter spacetime.

\smallskip

Furthermore, due to the special form of the metric, components of the Ricci tensor, as well as its trace-free part, are diagonal, so that
\be
\mathscr{R}_p = (g^{tt} R_{tt})^p + \sum_i (g^{ii} R_{ii})^p \quad \textrm{and} \quad \mathscr{S}_p = (g^{tt} S_{tt})^p + \sum_i (g^{ii} S_{ii})^p \, ,
\ee
and inequality (\ref{ineq:2m}) implies
\be
(\mathscr{R}_s)^{2m} \le D^{2m-s} (\mathscr{R}_{2m})^s \quad \textrm{and} \quad (\mathscr{S}_s)^{2m} \le D^{2m-s} (\mathscr{S}_{2m})^s
\ee
for all integers $s$ and $m$, such that $1 \le s < 2m$. If the metric is a solution of Einstein's gravitational field equation $G_{ab} = 8\pi T_{ab}$, then it immediately follows that the trace-free part of the energy-momentum tensor satisfies constraints (\ref{ineq:hatT}) irrespectively of its specific form.

\smallskip

To build some intuition, let us briefly look at more concrete examples of the metric function $f$ in $(1+3)$-dimensional spacetime. First of all, one might change the power of the radial coordinate in the Schwarzschild's $f_{\mathrm{Schw}}(r) = 1 - 2M r^{-1}$, as shown in Table \ref{tab:sphf} (``lower triangular'' population of unbounded invariants in the table is just a manifestation of the ``$R \to S \to K$'' rule). Apart from trivial $f(r) = 1$, corresponding to Minkowski spacetime, which renders all three basic invariants $R$, $S$ and $K$ bounded, the same properties are found among numerous regular black hole solutions \cite{Maeda22,SZ22,LYGM23} (examples with bounded basic curvature invariants, but diverging higher derivative invariants were studied by Musgrave and Lake \cite{ML95}). Also, one must bear in mind that it is possible to have bounded curvature invariants which nevertheless have pathological behaviour as in the example
\be
f(r) = 1 + e^{-r/a} (r/a)^n \sin(a/r) \, , \quad a > 0 \, ,
\ee
whose invariants ``wildly oscillate'' for $n=4$ in the limit $r \to 0^+$. Exponents $n < 4$ result in unbounded oscillating invariants as $r \to 0^+$, while for $n > 4$ these oscillations are ``damped''. A somewhat related phenomenon of ``wildly oscillating'' Kretschmann invariant in rotating black hole solutions was discussed by Ori \cite{Ori99}. Example $f(r) = 1 - C r^{-2}$, with bounded $R$ but divergent $S$ and $K$, is just a massless, $M=0$, member of the Reissner--Nordstr\"om family of solutions.

\bc
\begin{table}[ht]
\centering
\begin{tabular}{r|lll}
$\alpha$ & $R$ & $S$ & $K$ \\[0.2em]
$0$ & \checkmark & \checkmark & \checkmark \\
$1$ & \checkmark & \checkmark & \xmark \\
$2$ & \checkmark & \xmark & \xmark \\
$3$ & \xmark & \xmark & \xmark
\end{tabular}

\caption{Sample of examples in which basic curvature invariants either stay bounded, marked with \checkmark, or are unbounded, marked with \xmark, as $r \to 0^+$ for the $(1+3)$-dimensional static, spherically symmetric metric with $f(r) = 1 - C (1 - \delta_{\alpha 0}) r^{-\alpha}$ and constant $C \ne 0$.}\label{tab:sphf}
\end{table}
\ec

\smallskip

Higher derivative invariants for $(1+3)$-dimensional static, spherically symmetric spacetimes have the form
\begin{align}
\mathscr{D}_1 & = \frac{f}{r^6} \, \left ( r^3 f''' + 4r^2 f'' -2rf' + 4\left ( 1-f \right ) \right )^2 \, ,\\
\mathscr{D}_2 & = \frac{f}{r^6} \, \left ( 2r^2 \, \left ( \frac{1}{2} \, r^2 f''' + rf'' - f' \right )^2 + 12 \, \left ( \frac{1}{2} \, r^2 f'' + 1 - f \right )^2  \right ) \, ,\\
\mathscr{D}_3 & = \frac{f}{r^6} \, \left ( r^6 \left (f'''\right )^2 + 8r^2 \left ( r f'' - f' \right )^2 + 32 \, \left ( \frac{1}{2} \, rf' + 1 - f \right )^2  \right ) \, .
\end{align}
We can gain insight into the inequalities (\ref{D1D2}) and (\ref{D2D3}) from theorem \ref{tm:ineqD}, by writing differences between the left and right-hand side as products of $f(r)$ and manifestly nonnegative functions,
\begin{align}
4\mathscr{D}_2 - \mathscr{D}_1 & = \frac{f}{r^6} \, \left ( \left ( r^3 f''' - 2rf' - 4 + 4f \right )^2 + 4\left ( r^2 f'' + 2 - 2f \right )^2 \right ) \, ,\\
3\mathscr{D}_3 - 2\mathscr{D}_2 &= \frac{f}{r^6} \, \Big ( \left (r^3 f'''\right )^2 + \left ( r^3 f''' - 2r^2 f'' +  2rf'  \right )^2 + 8 \left (r^2 f''-r f' \right )^2 + \nonumber\\
& \hspace{50pt} + 2\left ( r^2 f'' - 4 r f' + 6f - 6 \right )^2 \Big ) \, .
\end{align}
From these expressions, it can be seen why the assumption $f>0$ in theorem \ref{tm:ineqD} is crucial. As a small addendum, we may illustrate the usage of the higher derivative invariants as horizon-detecting descriptors, as mentioned in the Introduction. For example, subextremal Reissner--Nordstr\"om black hole, with mass $M > 0$ and charge $Q$ satisfying $|Q| < M$, has metric function
\be
f(r) = 1 - \frac{2M}{r} + \frac{Q^2}{r^2} = \frac{1}{r^2} \, (r-r_-)(r-r_+) \, ,
\ee
with horizon radii $r_{\pm} = M \pm \sqrt{M^2-Q^2}$, and the corresponding invariants
\be
\mathscr{D}_1 = 0 \qqd \mathscr{D}_2 = \frac{80 Q^4}{r^{10}} \, f(r) \qqd \mathscr{D}_3 = \frac{16}{r^{10}} \, f(r) \left ( 45 M^2 r^2 - 108 M Q^2 r + 76 Q^4 \right ) \, .
\ee
We see that $\mathscr{D}_2$ and $\mathscr{D}_3$ have zeros exactly on horizons $r = r_\pm$; they are strictly positive for $r \in \left< 0,r_- \right> \cup \left< r_+,\infty \right>$ and strictly negative for $r \in \left< r_-,r_+ \right>$. This is not a universal feature, as Kerr's black hole invariant $\mathscr{D}_3$ changes sign not at the horizon, but on the ergosurface \cite{Gass98,Moffat14,Saa07} (in other words, $\mathscr{D}_3$ cannot be used as a reliable horizon detector in the non spherically symmetric case).

\subsection{No-go theorems for regular black holes with NLE fields} 

For many static, spherically symmetric spacetimes one can, in principle, find a corresponding NLE theory in which the metric is a solution of the corresponding field equations. Such reverse engineering in the NLE context, motivated by the quest for regular black hole solutions, was introduced by Bronnikov \cite{Bronnikov00,Bronnikov17}, rediscovered by Fan and Wang \cite{FW16} (cf.~Bronnikov's objections in \cite{Bronnikov17c}) and upgraded more recently in \cite{BFJS24}. Nevertheless, construction of regular black holes with NLE fields is constrained by Bronnikov's theorem \cite{Bronnikov00} for Lagrangians of the form $\LL(\FF)$ and more general theorems \cite{BSJ22a,BSJ22b} for Lagrangians of the form $\LL(\FF,\GG)$ (see an overview in \cite{Bronnikov22}). 

\smallskip

The proof strategy for these no-go theorems follows three main steps: (1) assume that some curvature invariants are bounded in the interior of the black hole, (2) use this assumption in various contractions of the gravitational field equation $E_{ab} = 8\pi T_{ab}$ to conclude that the corresponding contractions of the energy-momentum tensor are bounded, (3) use previous conclusion in the electromagnetic field equations and find an inconsistency with some general assumptions about the theory under consideration. Theorems in \cite{BSJ22a} were focused on $(1+3)$-dimensional Einstein's gravitational field equation, $E_{ab} = R_{ab} - \frac{1}{2} R g_{ab} + \Lambda g_{ab}$, in which the assumption about bounded $R$ and $S$, via contractions $g^{ab} E_{ab} = -R + 4\Lambda$ and $E_{ab} E^{ab} = S + 2\Lambda(2\Lambda - R)$, imply boundedness of $T$ and $T_{ab} T^{ab}$. Then, at least for the static spherically symmetric solutions, it was shown that this leads to an inconsistency with the Maxwellian weak field limit\footnote{We say that a NLE Lagrangian $\LL(\FF,\GG)$ of $C^1$ class obeys the Maxwellian weak field limit if $\dd_\FF \LL = -\frac{1}{4} + O(\mathcal{H})$ and $\dd_\GG \LL = O(\mathcal{H})$ as $\mathcal{H} \to 0$, where $\mathcal{H} = \sqrt{\FF^2 + \GG^2}$, on some neighbourhood of the $\FF$-$\GG$ plane.} of the NLE Lagrangian. In other words, given that we assume that NLE Lagrangian obeys Maxwellian weak field limit, then at least one of the invariants, $R$ or $S$, has to be unbounded as $r \to 0$. The conclusion is rather general if the black hole is strictly electrically charged, but relies on further Lagrangian assumptions in the presence of the magnetic charge (classification of the NLE Lagrangians which do or do not admit regular black hole solutions is still incomplete). All these NLE no-go theorems are strengthened with results of the current paper: ``$R \to S \to K$'' rule proves that in any case when either $R$ or $S$ is unbounded, Kretschmann's invariant $K$ will also necessarily be unbounded.

\smallskip

We may also look at further generalization in the family of $F(R)$-theories \cite{DFT,SF}, when the gravitational part of the field equation is defined by the $F(R)$ term in the Lagrangian, such that
\be
E_{ab} = F'(R) R_{ab} + \Big( \Box F'(R) - \frac{1}{2} \, F(R) \Big) g_{ab} - \nab{a} \nab{b} F'(R) .
\ee
Here we have
\begin{align}
g^{ab} E_{ab} & = -2F(R) + R F'(R) + 3\Box F'(R) , \\
E^{ab} E_{ab} & = S F'(R)^2 - 2F'(R) R^{ab} \nab{a}\nab{b} F'(R) + (\nab{a}\nab{b} F'(R))(\nabla^a \nabla^b F'(R)) + \nonumber\\
 & \hspace{10pt} +\big( 2\Box F'(R) - F(R) \big) \big( RF'(R) + \Box F'(R) - F(R) \big) .
\end{align}
Now, taking into account that
\begin{align*}
\nab{a} F'(R) & = F''(R) \nab{a} R , \\
\nab{a} \nab{b} F'(R) & = F'''(R) (\nab{a} R)(\nab{b} R) + F''(R) \nab{a} \nab{b} R ,
\end{align*}
one can see that $g^{ab} E_{ab}$ and $E^{ab} E_{ab}$ are polynomials constructed from derivatives $F^{(n)}(R)$ and elements of the set of invariants 
\begin{align}
\mathcal{J} & = \big\{ R, S, (\nab{a} R)(\nabla^a R), R^{ab}(\nab{a} R)(\nab{b} R), R^{ab}(\nab{a} \nab{b} R), \Box R, \nonumber\\
& \hspace{30pt} (\nabla^a \nabla^b R)(\nab{a}\nab{b} R), (\nabla^a \nabla^b R)(\nab{a} R)(\nab{b} R) \big\} .
\end{align}
Hence, given that one assumes that all the invariants from the set $\mathcal{J}$ are bounded in the interior of static spherically symmetric black hole (it is sufficient to assume that they are continuous on a closed ball $0 \le r \le r_0$ for some $r_0 > 0$ and any fixed $t$), as well as $F(R)$ and its derivatives $F'(R)$, $F''(R)$ and $F'''(R)$ on the range of the Ricci scalar $R$ (it is enough to assume that function $F$ is of class $C^3$ on the image of the closed ball $0 \le r \le r_0$ under Ricci scalar $R$), one can repeat the proof strategy described above. The conclusion in this generalization will be inconsistency between the Maxwellian weak field limit of the NLE Lagrangian and at least some of the assumed regularity conditions of the curvature invariants. In principle, these assumptions may be optimized and conclusions strengthened, provided that one finds additional inequalities between the invariants from the set $\mathcal{J}$, but we leave this line of research for future work.

\subsection{FLRW spacetimes} 

Friedmann--Lema\^itre--Robertson--Walker (FLRW) spacetimes are spatially homogeneous and isotropic, with the metric
\be
\df s^2 = -\df t^2 + a(t)^2 \left ( \frac{\df r^2}{1 - kr^2} + r^2 \left ( \df\theta^2 + \sin^2\theta \, \df\varphi^2 \right ) \right ) \, .
\ee
Function $a(t)$ is known as the scale parameter and parameter $k \in \{0,\pm 1\}$ distinguishes between the cases of flat ($k=0$), positively ($k = 1$) and negatively ($k = -1$) curved $t=\textrm{const}.$ hypersurfaces. Basic curvature invariants for the FLRW metric read
\begin{align}
R & = \frac{6}{a^2} \, (a \ddot{a} + \dot{a}^2 + k) \, , \\
S & = \frac{12}{a^4} \, \left ( (a\ddot{a})^2 + a\ddot{a}(\dot{a}^2 + k) + (\dot{a}^2 + k)^2 \right ) \, , \\
K & = \frac{12}{a^4} \, \left ( (a\ddot{a})^2 + (\dot{a}^2 + k)^2 \right ) \, .
\end{align}
As the Weyl tensor is identically zero (FLRW metric belongs to the Petrov type O), we have $W = 0$ and
\be
4S - R^2 = 3K - 2S = \frac{12}{a^4} \, \left ( a\ddot{a} - \dot{a}^2 - k \right )^2 \, ,
\ee
in accordance with the equation (\ref{eq:RSKWb}).

\bc
\begin{table}[ht]
\centering
\begin{tabular}{r|lll}
$w$ & $R$ & $S$ & $K$ \\[0.3em]
$-1$ & \checkmark & \checkmark & \checkmark \\
impossible & \checkmark & \checkmark & \xmark \\
$1/3$ & \checkmark & \xmark & \xmark \\
$\notin \{-1, 1/3\}$ & \xmark & \xmark & \xmark
\end{tabular}

\caption{Sample of examples in which basic curvature invariants either stay bounded, marked with \checkmark, or are unbounded, marked with \xmark, as $t \to 0^+$ for the FLRW metric.}\label{tab:FLRW}
\end{table}
\ec

Again, for more concrete examples we look at the ideal fluid, with a linear equation of state $p = w\rho$, in flat FLRW spacetime. Solutions of the Friedmann equations are
\be
\frac{a(t)}{a_0} = \left\{ \begin{array}{rl} t^{\frac{2}{3(1+w)}} \, , & w \ne -1 \\ \exp(Ct) \, , & w = -1 \end{array} \right.
\ee
with constants $a_0 > 0$ and $C \ne 0$. The $w=-1$ case is essentially (anti)-de Sitter space with $C = \sqrt{|\Lambda|/3}$. Corresponding basic curvature invariants for $w \ne -1$ are
\be
R = \frac{4}{3} \, \frac{1-3w}{(1+w)^2 t^2} \qqd S = \frac{16}{9} \, \frac{1+3w^2}{(1+w)^4 t^4} \qqd K = \frac{16}{27} \, \frac{4 + (1+3w)^2}{(1+w)^4 t^4} \, ,
\ee
related via
\be
4S - R^2 = 3K - 2S = \frac{16}{3(1+w)^2 t^4} \, .
\ee
In the $w = -1$ case, we have simply $R = 12 C^2$, $S = 36 C^4$, $K = 24 C^4$ and $4S - R^2 = 3K - 2S = 0$. Apart from the special $w = 1/3$ case (used to model the radiation-dominated era of the universe), with $a(t) = a_0 \sqrt{t}$, and $w = -1$ case (e.g.~de Sitter space with $\rho = -p = \Lambda/8\pi > 0$), all other values of parameter $w$ lead to having solutions in which all three invariants $R$, $S$ and $K$ diverge as $t \to 0^+$ (see table \ref{tab:FLRW}). The case with bounded $R$ and $S$, but unbounded $K$ is excluded by the relation $3K = 6S - R^2$.

\subsection{Bianchi type I} 

Homogeneous anisotropic cosmological models are based on Bianchi's classification. The simplest example of Bianchi universes is known as Bianchi type I, which is an immediate generalization of the flat FLRW metric to the case where each spatial direction has a different scale factor in general. The general form of the metric for Bianchi type I is
\be
\df s^2 = - \df t^2 + \sum_{i=1}^{D-1} a_i^2(t) (\df x^i)^2 \, .
\ee
Three basic curvature invariants have simple forms
\begin{align}
R & = \sum_i \left ( 2\,\frac{\ddot{a}_i}{a_i} + \sum_{j\ne i} \frac{\dot{a}_i \dot{a}_j}{a_i a_j} \right ) , \\
S & = \left ( \sum_i \frac{\ddot{a}_i}{a_i} \right )^{\!2} + \sum_i \left ( \frac{\ddot{a}_i}{a_i} + \sum_{j\ne i} \frac{\dot{a}_i \dot{a}_j}{a_i a_j} \right )^{\!2} , \\
K & = 4 \sum_i \frac{\ddot{a}_i^2}{a_i^2} + 2 \sum_i \sum_{j\ne i} \frac{(\dot{a}_i \dot{a}_j)^2}{(a_i a_j)^2} \, ,
\end{align}
where all sums are assumed to go from $1$ to $D-1$. Using abbreviations
\be
b_i \defeq \frac{\ddot{a}_i}{a_i} + \sum_{j\ne i} \frac{\dot{a}_i \dot{a}_j}{a_i a_j} \qqd b_D \defeq \sum_i \frac{\ddot{a}_i}{a_i} \, ,
\ee
we can write
\be
DS - R^2 = D \sum_{i=1}^D b_i^2 - \left ( \sum_{i=1}^D b_i \right )^{\!2} = \frac{1}{2} \sum_{i,j=1}^D (b_i - b_j)^2 \ge 0 \, .
\ee
which is a manifestly nonnegative expression that proves inequality (\ref{ineq:RS}). In the same way, using abbreviations
\be
c_i \defeq \frac{\ddot{a}_i}{a_i} \qqd d_{ij} \defeq \frac{\dot{a}_i \dot{a}_j}{a_i a_j} \, ,
\ee
and applying the AM-QM inequality, we can write
\begin{align}
\frac{D-1}{2} \, K - S & = (D-1) \sum_i c_i^2 - \left ( \sum_i c_i \right )^2 + \nonumber \\
 & \hspace{20pt} + \sum_i \left ( (D-1) \left ( c_i^2 + \sum_{j\ne i} d_{ij}^2 \right ) - \left ( c_i + \sum_{j\ne i} d_{ij} \right )^2 \right ) \ge 0
\end{align}
which proves inequality (\ref{ineq:SK}).

\smallskip

Let us now look at the examples of scale factors in $(1+3)$-dimensional Bianchi type I spacetime and the associated basic curvature invariants, shown in the table \ref{tab:Bianchi}. If all three scale factors $(a_1,a_2,a_3)$ are exponential functions then $R$, $S$ and $K$ remain bounded as $t \to 0^+$. Examples of other behaviour of curvature invariants are most easily constructed with power functions of the form $a_i(t) = (t/t_0)^{e_i}$, constant $t_0 > 0$ and real exponents $(e_1,e_2,e_3)$. In this case we have $R = 2 P_R(e_1,e_2,e_3) t^{-2}$, $S = 2 P_S(e_1,e_2,e_3) t^{-4}$ and $K = 4 P_K(e_1,e_2,e_3) t^{-4}$ with polynomials
\begin{align}
P_R(e_1,e_2,e_3) & = e_1^2 + e_2^2 + e_3^2 - e_1 - e_2 - e_3 + e_1 e_2 + e_2 e_3 + e_3 e_1 \, ,\\
P_S(e_1,e_2,e_3) & = (e_1^2 + e_2^2 + e_3^2 - e_1 - e_2 - e_3)^2 \, + \nonumber\\
 & \hspace{70pt} + (e_1^2 + e_2^2 + e_3^2 - 1) (e_1 e_2 + e_2 e_3 + e_3 e_1) \, ,\\
P_K(e_1,e_2,e_3) & = e_1^2 (e_1-1)^2 + e_2^2 (e_2-1)^2 + e_3^2 (e_3-1)^2 + e_1^2 e_2^2 + e_2^2 e_3^2 + e_3^2 e_1^2 \, .
\end{align}
In order to have bounded $R$ and $S$ as $t \to 0^+$ we demand $P_R = 0$ and $P_S = 0$, which combined imply
\be
(e_1 e_2 + e_1 e_3 + e_2 e_3)(e_1 + e_2 + e_3 - 1) = 0 \, .
\ee
If $e_1 e_2 + e_2 e_3 + e_3 e_1 = 0$, condition $P_R = 0$ implies $e_1^2 + e_2^2 + e_3^2 = e_1 + e_2 + e_3$ (and, consequently, $P_S = 0$). In other words, we have $(e_1 + e_2 + e_3)^2 = e_1 + e_2 + e_3$, which is possible iff $e_1 + e_2 + e_3 = 0$ (which leads to the trivial case) or $e_1 + e_2 + e_3 = 1$. On the other hand, if $e_1 + e_2 + e_3 = 1$, taking into account that $P_R = (e_1 + e_2 + e_3)(e_1 + e_2 + e_3 - 1) - (e_1 e_2 + e_2 e_3 + e_1 e_3)$, condition $P_R = 0$ implies $e_1 e_2 + e_2 e_3 + e_3 e_1 = 0$, and then from $P_S = 0$ we get $e_1^2 + e_2^2 + e_3^2 = 1$. In conclusion, for $(e_1,e_2,e_3) \ne (0,0,0)$ we get the system
\begin{align}
e_1 + e_2 + e_3 & = 1 \, ,\\
e_1^2 + e_2^2 + e_3^2 & = 1 \, ,\\
e_1 e_2 + e_2 e_3 + e_3 e_1 & = 0 \, ,
\end{align}
which can be solved in the form
\be
e_2 = \frac{1}{2} \, \left ( 1 - e_1 \pm \sqrt{(1-e_1)(1+3e_1)} \right ) , \quad e_3 = 1 - e_1 - e_2 \, .
\ee
If any of the exponents is trivial, e.g.~$e_1 = 0$, then the only solutions of the system are $(e_1,e_2,e_3) = (0,1,0)$ and $(e_1,e_2,e_3) = (0,0,1)$. Recent complementary analyses of singularities in cosmological models may be found in \cite{NY21} and \cite{CKMP23}.

\bc
\begin{table}[ht]
\centering
\begin{tabular}{rrr|lll}
$a_1(t)$ & $a_2(t)$ & $a_3(t)$ & $R$ & $S$ & $K$ \\[0.2em]
$e^{\alpha t}$ & $e^{\beta t}$ & $e^{\gamma t}$ & \checkmark & \checkmark & \checkmark \\
$t^{-\frac{1}{3}}$ & $t^{\frac{2}{3}}$ & $t^{\frac{2}{3}}$ & \checkmark & \checkmark & \xmark \\
$t^{\frac{1}{2}}$ & $t^{-\frac{1}{6}}$ & $t^{-\frac{1}{6}}$ & \checkmark & \xmark & \xmark \\
$t^{-\frac{1}{2}}$ & $t^{-\frac{1}{6}}$ & $t^{-\frac{1}{6}}$ & \xmark & \xmark & \xmark
\end{tabular}

\caption{Sample of examples in which basic curvature invariants either stay bounded, marked with \checkmark, or are unbounded, marked with \xmark, as $t \to 0^+$ for the Bianchi type I metric. Coefficients $\alpha$, $\beta$ and $\gamma$ are arbitrary real numbers.}\label{tab:Bianchi}
\end{table}
\ec

\subsection{Schmidt's spacetime} 

It is interesting to see how even a ``slight'' deviation from the family of spacetimes in focus provides a counterexample to inequalities. Schmidt \cite{Schmidt03} has introduced a metric
\be\label{eq:Schmidt}
\df s^2 = -\df t^2 + 2yz \, \df t \, \df x + \df x^2 + \df y^2 + \df z^2
\ee
on $\rr^4$ in order to demonstrate that the ``Weyl squared'' invariant $W$ is not necessarily positive definite. Namely, here we have
\be
W(y,z) = \frac{(y^2 z^2 - 3)(y^2 + z^2)^2 y^2 z^2 + 12(y^4 z^4 - 1)}{3(1 + y^2 z^2)^4} \, , 
\ee
so that e.g.~$W(0,0) = -4 < 0$ and $W(1,3) = 53/250 > 0$. Moreover, Schmidt's metric also breaks other inequalities. Differences $\delta_1 \defeq 4S - R^2$ and $\delta_2 \defeq 3K - 2S$ for (\ref{eq:Schmidt}) are given by
\begin{align}
\delta_1(y,z) & = \frac{(3y^4 z^4 - 10 y^2 z^2 + 3)(y^2 + z^2)^2 + 64(1 + y^2 z^2)y^2 z^2}{4 (1 + y^2 z^2)^4} \, , \\
\delta_2(y,z) & = \frac{(7y^4 z^4 - 22y^2 z^2 + 3)(y^2 + z^2)^2 + 16(7y^4 z^4 + 4y^2 z^2 - 3)}{4 (1 + y^2 z^2)^4} \, .
\end{align}
As $\delta_1(0,3) = 243/4 > 0$ and $\delta_2(0,3) = 195/4 > 0$, but $\delta_1(1/3,3) = -1033/324 < 0$ and $\delta_2(1/3,3) = -1465/108 < 0$, neither $\delta_1$ nor $\delta_2$ is (positive or negative) definite. This is not a surprise, as Schmidt's metric does not belong to any of the families of spacetimes discussed above: although $\dd_y$ and $\dd_z$ are hypersurface orthogonal vector fields, $t$-$x$ metric block is $yz$-dependent.

\section{Final remarks} \label{sec:final} 

Inequalities between the curvature invariants, which have been proven and analyzed in the paper, are the first step towards a better understanding of relations between the invariants and their role in the classification of spacetime singularities. Ricci scalar $R$, ``Ricci squared'' $S$ and Kretschmann invariant $K$ are constrained by the theorem \ref{tm:ineqRSK} (also, invariants $\mathscr{D}_1$, $\mathscr{D}_2$ and $\mathscr{D}_3$ by the theorem \ref{tm:ineqD}), independently of the gravitational field equations. Further constraints via energy-momentum tensor are given by the theorems \ref{tm:Tineq} and \ref{tm:TNLE} for the solutions of the Einstein's gravitational field equation, independently of the specific form of the spacetime metric $g_{ab}$. All these results were illustrated and discussed with numerous spacetime examples in section \ref{sec:examples}. 

\smallskip

An immediate question may be posed about the geometric characterization of the family of metrics used in the theorem \ref{tm:ineqRSK}. In any of the three types of the metric, vector field $X^a = (\dd_t)^a$ is hypersurface orthogonal, which is reflected in the block-diagonal form of the metric \cite{BS23}, but it is not necessarily a Killing vector field. Crucial detail, vanishing of the Riemann components $R_{tijk} = 0$, captured by the lemma \ref{lm:RCzero}, may be translated into condition $(\nab{i}\nab{j} - \nab{j}\nab{i}) X_k = 0$ for all $i,j,k \in \{1,\dots,D-1\}$. It would be interesting to find the most general spacetime metric satisfying this condition.

\smallskip

It is well known that different notions of singular spacetimes, those based on inextendible incomplete geodesics and those based on unbounded curvature, are \emph{generally} independent, with various examples which are singular in one sense, regular in other and vice versa \cite{Geroch68,ES77,ES79,CGL83}. This, however, does not imply that geodesic and curvature singularities might not be connected for some \emph{special} families of spacetimes (which could be physically relevant). Indeed, Clark \cite{Clarke75} and Tipler \cite{Tipler77} have proven theorems which bound the growth of components of Riemann and Ricci tensors along geodesics, followed by a series of related results \cite{Szabados82,Newman84,Szabados87,KR92,KL95,Racz23}. We believe that further study of inequalities among the curvature invariants might prove relevant for a deeper understanding of classical spacetime singularities, their quantum resolution in the path integral approach \cite{BE21,Borissova24} and possibly even Penrose's Weyl curvature hypothesis \cite{Penrose77,GH02}.

\section*{Acknowledgements}
The research was supported by the Croatian Science Foundation Project No.~IP-2020-02-9614.


\appendix 

\section{Brief overview of nonlinear electromagnetism} \label{sec:NLE} 

Nonlinear electromagnetism (NLE) is an umbrella term for various nonlinear extensions of classical Maxwell's electromagnetism, motivated either by the quantum corrections or introduced by hand as phenomenological attempts to tame singularities of the classical theory \cite{Plebanski70,Sorokin22}. We shall use the term in a slightly narrow sense, for any theory with the Lagrangian density $\LL(\FF,\GG)$ which is a function of two electromagnetic invariants $\FF \defeq F_{ab} F^{ab}$ and $\GG \defeq F_{ab} {\hdg F}^{ab}$. Corresponding generalized source-free Maxwell's equations (cf.~\cite{BSJ21}) read $\df \F = 0$ and $\df{\hdg(} \LL_\FF \F + \LL_\GG {\hdg\F}) = 0$, with the abbreviations $\LL_\FF \defeq \dd_\FF \LL$ and $\LL_\GG \defeq \dd_\GG \LL$. Furthermore, NLE energy-momentum tensor may be written in the form
\be
T_{ab} = -4\LL_\FF T^{\mathrm{(Max)}}_{ab} + \frac{1}{4} \, T g_{ab}
\ee
with trace $T \defeq g^{ab} T_{ab}$ and Maxwell's electromagnetic energy-momentum tensor
\be
T^{\mathrm{(Max)}}_{ab} = \frac{1}{4\pi} \left( F_{ac} \tensor{F}{_b^c} - \frac{1}{4} \, \FF g_{ab} \right) .
\ee
Hence, the corresponding trace-free part of the NLE energy-momentum tensor reduces to
\be\label{eq:trfreeNLE}
\widehat{T}_{ab} = -4\LL_\FF T_{ab}^{\mathrm{(Max)}} \, .
\ee
NLE field satisfies the null energy condition if and only if $\LL_\FF \le 0$ and it satisfies the dominant energy condition if and only if $\LL_\FF \le 0$ and $T \le 0$ (cf.~discussions in \cite{Plebanski70} and \cite{BSJ21}). One of the curiosities of the NLE theories is so-called stealth fields, those for which the energy-momentum tensor vanishes $T_{ab} = 0$ although the field itself is nontrivial, $F_{ab} \ne 0$ (hence, stealth fields do not gravitate). It is known \cite{ISm17} that stealth fields are absent in Maxwell's theory, but occur in a NLE theory if and only if $\LL_\FF = 0$ and $T = 0$ at the given point (in other words, stealth NLE fields ``live'' on the borderline of energy conditions).

\section{ZM invariants for (1+3)-dimensional static, spherically symmetric spacetimes} \label{sec:ZM} 

The set of Zakhary--McIntosh (ZM) invariants consists of 17 scalars and it is the smallest known set of polynomial curvature invariants necessarily containing a maximal set of algebraically independent scalars in a $(1+3)$-dimensional spacetime. The ZM invariants are sufficient to distinguish between all Petrov and Segre types of a metric \cite{CMc91,ZMc97}.

\smallskip

We will use expressions for ZM invariants from \cite{OCWH20} (Eq.~(10)) with slight modifications. Since the invariants $I_{13}$ and $I_{14}$ have not been explicitly written in that paper, we use
\begin{align}
I_{13} &= R_{ab} \, R^{bc} \, R_{de} \, R^{ef} \, \tensor{C}{^a^d_c_f} \, ,\\
I_{14} &= - R_{ab} \, R^{bc} \, R_{de} \, R^{ef} \, \tensor{{\hdg C}}{^a^d_c_f} \, .
\end{align}
We also provide a corrected formula for $I_{16}$ due to a possible typo in \cite{OCWH20}
\begin{align}
I_{16} &= -\frac{1}{32} \, R_{ab} \, R_{cd} \, \Big ( C_{eghf} \, C^{eabf} \, C^{gcdh} + C_{eghf} \, {\hdg C}^{eabf} \, {\hdg C}^{gcdh} - \nonumber \\
 & \hspace{90pt} - {\hdg C}_{eghf} \, {\hdg C}^{eabf} \, C^{gcdh} + {\hdg C}_{eghf} \, C^{eabf} \, {\hdg C}^{gcdh} \Big ) \, .
\end{align}
Let us turn to a $(1+3)$-dimensional static, spherically symmetric spacetime, described by the metric
\be \label{eq:staticSphericallySymmetric}
\df s^2 = -f(r) \, \df t^2 + \frac{\df r^2}{f(r)} + r^2 \df\Omega_{2}^2 \, .
\ee
Using the abbreviations
\begin{align}
    g(r) & = f'' -\frac{2}{r} f' + \frac{2}{r^2} \, (f-1) \label{eq:abb1} \\
    h(r) & = f'' +\frac{4}{r} f' + \frac{2}{r^2} \, (f-1) \label{eq:abb2} \\
    u(r) & = \frac{1}{2} f'' + \frac{f'}{r} \label{eq:abb3} \\
    v(r) & = \frac{f'}{r} + \frac {f-1}{r^2} \label{eq:abb4} \\
    z(r) & = f'' - \frac{2}{r^2} \, (f-1) \label{eq:abb5}
\end{align}
explicit expressions for the corresponding ZM invariants are summarized in table \ref{tab:ZMsph}. This list can be helpful in further investigation of inequalities among the invariants in general static spacetimes.

\bc
\begin{table}[ht!]
\centering
\arrs{1.2}
\begin{tabular}{lll}
Weyl invariants \hspace{7pt} & Ricci invariants \hspace{7pt} & Mixed invariants \\[0.1em]
$I_1 = \frac{1}{3} \, g^2$ & $I_5 = - h$ & $I_9 = \frac{1}{12} \, z^2 g$ \\
$I_2 = 0$ & $I_6 = 2 u^2 + 2 v^2$ & $I_{10} = 0$ \\
$I_3 = -\frac{1}{18} \, g^3$ & $I_7 = -2 u^3 - 2 v^3$ & $I_{11} = \frac{1}{36} \, z^2 g^2$ \\
$I_4 = 0$ & $I_8 = 2u^4 + 2 v^4$ & $I_{12} = 0$ \\
 & & $I_{13} = -\frac{1}{48} \, z^2 g h^2$ \\
 & & $I_{14} = 0$ \\
 & & $I_{15} = \frac{1}{576} \, z^2 g^2$ \\
 & & $I_{16} = -\frac{1}{3456} \, z^2 g^3$ \\
 & & $I_{17} = 0$
\end{tabular}
\arrs{1.0}

\caption{ZM invariants for the metric (\ref{eq:staticSphericallySymmetric}) with the abbreviations (\ref{eq:abb1})--(\ref{eq:abb5}).} \label{tab:ZMsph}
\end{table}
\ec

\bibliographystyle{amsalpha}
\bibliography{wci}

\end{document}